# A photo-click thiol-ene collagen-based hydrogel platform for skeletal muscle tissue engineering

*Roisin Holmes, Xuebin B. Yang, David J. Wood, and Giuseppe Tronci\**

R. Holmes, X.B. Yang, D.J. Wood, G. Tronci
Biomaterials and Tissue Engineering Research Group, School of Dentistry, Wellcome Trust Brenner Building, St. James's University Hospital, University of Leeds, Leeds, LS9 7TF, United Kingdom
E-mail: g.tronci@leeds.ac.uk

R. Holmes
Institute of Medical and Biological Engineering, School of Mechanical Engineering, University of Leeds, Leeds LS2 9JT, UK

G. Tronci
Clothworkers' Centre for Textile Materials Innovation for Healthcare, School of Design, University of Leeds, Leeds, LS2 9JT, United Kingdom

Keywords: photo-click, hydrogels, collagen, biocompatibility, skeletal muscle regeneration

**Abstract**
UV-cured collagen-based hydrogels hold promise in skeletal muscle regeneration due to their soft elastic properties and porous architecture. However, the complex triple helix conformation of collagen and environmental conditions, i.e. molecular oxygen, pose risks to reaction controllability, wet-state integrity and reproducibility. To address this challenge, a photo-click hydrogel platform is presented through an oxygen-insensitive thiol-ene reaction between 2-iminothiolane (2IT)-functionalized type I collagen and multi-arm, non-homopolymerizable norbornene-terminated polyethylene glycol (PEG). UV-induced network formation is demonstrated by oscillatory time sweeps on the reacting thiol-ene mixture, so that significantly increased storage moduli are measured and adjusted depending on the photo-initiator concentration. Variations in PEG functionality (4-arm and 8-arm) and PEG content generate hydrogels with skeletal muscle native stiffness ($E_c$= 1.3±0.2–11.5±0.9 kPa), diffusion-controlled swelling behavior and erosion-driven degradability. *In vitro*, no cytotoxic effect is detected on C2C12 murine myoblasts, while myogenic differentiation is successfully accomplished on hydrogel-seeded cells in low serum culture medium. *In vivo*, 7-day subcutaneous implantation of selected thiol-ene hydrogel in rats reveal higher cell infiltration, blood vessel formation, and denser tissue interface compared to a clinical gold standard collagen matrix (Mucograft®, Geistlich). These results therefore support the applicability and further development of this hydrogel platform for skeletal muscle regeneration.

**1. Introduction**

Skeletal muscle is one of the most abundant tissues in the human body and is responsible for locomotion, internal organ protection and metabolic regulation. Its native function and self-healing capability are hampered when more than 20% of the tissue is lost, yielding fibrous scar tissue of minimal functionality.[1] Current clinical treatments for volumetric muscle loss rely on the transplantation of biological grafts to restore tissue function, though issues with graft source availability, donor site morbidity and immunotoxicity make these approaches unsustainable in the long term.

A promising strategy to reverse this trend and accomplish skeletal muscle regeneration relies on the development of a regenerative device that can be implanted at the site of the injury to enable region-specific satellite cell activation and myoblast differentiation.[2] In this context, the device should be made of a biomaterial capable of developing a pro-healing milieu *in situ*, while being free of cellular and pharmacological factors, to ensure a regulatory-friendly and cost-effective translation pathway to the clinic.[3] A promising concept to realize such device-induced pro-healing milieu is to equip the biomaterial with specific biochemical and biophysical cues, aiming to promote satellite cell attachment, proliferation and cell fusion towards myofiber formation. Multiple prototypes have been realized at the lab scale, and tested *in vitro*, looking at the effect of substrate stiffness and nano-to-microscale organization.[4] Nevertheless, further development aiming to assess the prototype biocompatibility *in vivo* and compare it to clinical gold standards appears limited. These investigations are critical to enable clinical translation and meet the demands of end-users and manufacturers.

Hydrogels are water-swollen materials typically made of covalently crosslinked hydrophilic polymers of either synthetic or natural origin. Their high water content and mechanical softness mimic aspects of the extracellular matrix (ECM) of human tissues, making them appealing for a range of regenerative applications, including wound healing,[5] as well as bone,[6] cartilage,[7] and skeletal muscle[8] tissue engineering. In this context, the application of ECM-extracted biopolymers as hydrogel building block enables integration of inherent biochemical features, key to successfully accomplishing cell adhesion and differentiation, with water retention capability and device biodegradability.[9] These attributes are highly appealing, aiming to avoid additional functionalization steps at the hydrogel design stage, which are otherwise required when employing synthetic or natural polymers of other origin.[10]

A major biopolymer of the ECM is type I collagen, which is found in commercial wound dressings,[11] dental membranes,[12] and soft tissue matrices,[13] and whose molecular structure and nanoscale organization play a crucial role in myogenesis, wound healing and tissue remodeling. Type I collagen is found in a chemically crosslinked state *in vivo* and represents a major structural component of the basal lamina of muscle fibers. It presents cell-binding sequences, notably the RGD motif, which have been found to support the adhesion and differentiation of C2C12 murine myoblasts.[14] On the other hand, type I collagen extracted from biological tissues lacks its native crosslinked configuration found *in vivo*, resulting in poor mechanical properties, instability and uncontrollable degradation in physiological conditions. The presence of multiple functional groups on the collagen molecule allows for chemical accessibility, which offers a means to control the material stiffness, microstructure and macroscopic properties. However, modifying collagens is challenging because of the presence of both strong electrophilic and nucleophilic functional groups on the same backbone.[15] Although post-extraction synthetic strategies have been developed to restore chemical crosslinks in type I collagen *ex vivo*, they are still associated with controllability issues and environmental limitations, e.g. the presence of oxygen, restricting the wider applicability of type I collagen in regenerative devices.

Click chemistry is a class of environment-friendly reactions characterized by high reactivity and selectivity,[16] which has recently been employed in a range of biomedical applications, including cancer diagnostics,[17] controlled drug delivery,[18] micro-robots,[19] and tissue therapeutics.[20] These reactions generate covalent bonds in an oxygen-insensitive fashion with high yields and under mild reaction conditions, with no risks of side reactions or formation of potentially toxic by-products. Consequently, click chemistry is an appealing approach for the design of collagen hydrogels with defined structure-property relationships and a high degree of controllability, independent of the environmental conditions.

The photo-click thiol-ene reaction has emerged as an attractive click chemistry aiming to accomplish covalent thioether linkages *in situ* through a UV-induced, radical-mediated, step-growth mechanism involving a thiol and a vinyl functionality in the presence of a photo-initiator.[21] To ensure selectivity and controlled network architecture, electron-rich vinyl functionalities incapable of homo-polymerization should be employed, such as vinyl ethers or norbornene-based compounds.[16b] Although classic, UV-mediated crosslinking reactions have been successfully applied,[22] the photo-induced nature and cellular tolerability make thiol-ene reaction attractive for the realization of collagen-based regenerative systems with skeletal muscle native stiffness, enabling on-demand gelation, material stiffening, and biocompatibility *in vitro* and *in vivo*. The wide selectivity of this reaction can also be leveraged to overcome the poor enzymatic stability of type I collagen *in vivo* through crosslinking with synthetic macromers, e.g. polyethylene glycol (PEG). In contrast to the thiol-ene approach, covalent networks of type I collagen with PEG macromers have primarily been accomplished through crosslinking reactions that are susceptible to environmental conditions, e.g. molecular oxygen.[23] This susceptibility raises controllability issues with respect to macroscopic properties and material bio-functionality, hindering clinical translation.

Accordingly, this study aims to investigate the design of a new platform of photo-click thiol-ene hydrogels and assess the hydrogel applicability as a regenerative device for skeletal muscle applications. Our approach to thiol-ene network formation relies on the derivatization of type I collagen with 2-iminothiolane (2IT),[24] which was previously confirmed to enable more than 80 mol.% of lysine functionalization and a degree of triple helix retention of at least 90%. Either 4-arm (PEG4NB) or 8-arm (PEG8NB) norbornene-terminated PEG was employed as a non-homo-polymerizable macromer, whereby the impact of molecular architecture and crosslink density was studied on the physical properties of the thiol-ene hydrogel network and the response of C2C12 murine myoblasts. Further to systematic testing *in vitro*, a subcutaneous implantation study was also performed in rats, aiming to assess and compare the biocompatibility of the hydrogel *in vivo* with respect to that of a commercially available collagen-based matrix.

## 2. Results and discussion

The development of a thiol-ene collagen-based hydrogel platform is presented, aiming to assess the applicability of these hydrogels in skeletal muscle tissue regeneration according to the following characteristics: (1) on-demand network formation, (2) skeletal muscle-mimicking mechanical properties, (3) myoblast differentiation and (4) tissue tolerability. Type I rat tail collagen (RTC) was reacted with 2IT to accomplish a thiol-ene reaction-compliant terminal sulfhydryl group on lysine side chains (**Figure 1**), whilst maintaining the triple helix structure of native collagen molecule (**Table S1**). In accordance with the thiol-ene mechanism, the reaction proceeds through the addition of a thiyl radical to a vinyl generating a carbon-centered radical, which then abstracts a hydrogen from a thiol, yielding the thioether crosslinking bond and regenerating a thiyl radical. This mechanism was supported by control tests, whereby no UV-induced gelation was detected with LAP-supplemented aqueous solution containing either 2IT-functionalised rat tail collagen (RTC-2IT) or multi-arm norbornene-terminated PEG (PEGNB). As a non-homo-polymerizable vinyl-bearing macromer, the use of PEGNB, therefore, ensures a selective thiol-ene crosslinking reaction.

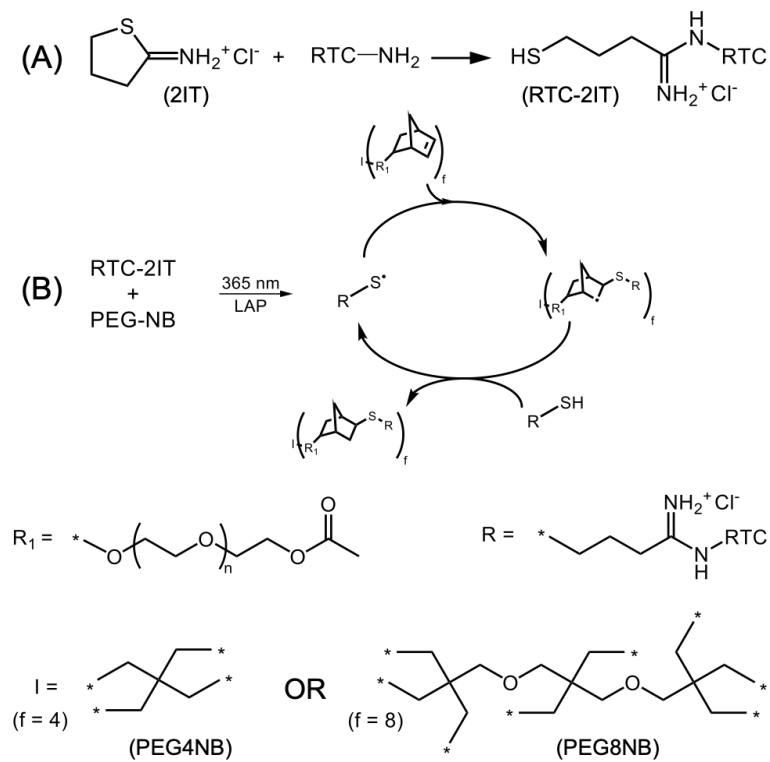

**Figure 1.** Design of photo-click thiol-ene hydrogels. (A) 2IT is reacted with RTC to generate a thiolated collagen derivative (RTC-2IT). (B) The thiolated product is dissolved in a LAP-supplemented aqueous solution containing either 8-arm (f=8, PEG8NB) or 4-arm (f=4, PEG4NB) norbornene-terminated PEG. UV exposure ($\lambda$: 365 nm) initiates the oxygen-insensitive thiol-ene reaction, yielding a covalently crosslinked hydrogel network.

## 2.1. Synthesis of thiol-ene hydrogels

The functionalization of RTC with 2IT proceeds via an amine-initiated ring-opening reaction, resulting in the derivatization of a primary amino group into a thiol (Figure 1a). Given the well-known limitations associated with the direct assessment of sulfyhydryl residues,[25] thiolation of the 2IT-reacted collagen product (RTC-2IT) was confirmed by quantifying the consumption of the primary amino groups through a 2,4,6-trinitrobenzenesulfonic acid (TNBS) assay. The resulting product revealed a significant degree of functionalization ($F = 72 \pm 2$ mol.%) and nearly retained triple helices ($X_c \sim 99\%$, Table S1). This value of $F$ was higher than the one measured in previously reported thiolated collagen,[15] and was comparable to the one of linear thiolated gelatin.[24] This observation indicates that the reaction with 2IT enabled high yield regardless of the complex triple helix conformation of type I collagen.

Having confirmed the synthesis of the thiolated collagen, attention moved to the investigation of the photo-click reaction. UV light was activated on a LAP-supplemented PBS solution containing the sample of RTC-2IT and PEG8NB ([NB]·[SH]$^{-1}$: 5.25). Complete gelation of reacting mixtures supplemented with either low (0.1% w/v) or high (0.5% w/v) concentration of LAP was observed via oscillatory time sweeps within 300 seconds of UV-curing, while no phase change was observed in the absence of UV irradiation (**Figure 2**a). The concentration of LAP was found to significantly affect the gelation kinetics and storage modulus ($G'$) of the formed hydrogel, confirming its direct effect on the amount of thiyl radicals generated. The thiol-ene mixture supplemented with high LAP (0.5 % w/v) concentration reached a $G'$ value of 100 Pa in less than 100 seconds of UV exposure, in contrast to the significantly longer time (180 seconds) required by the same reacting mixture with lower LAP (0.1 % w/v) concentration.

The major role played by the photo-initiator concentration on the gelation kinetics was also reflected by the final values of shear modulus, whereby a $G'$ of either 300 Pa or 4300 Pa was recorded with hydrogels prepared with low and high LAP concentration, respectively.

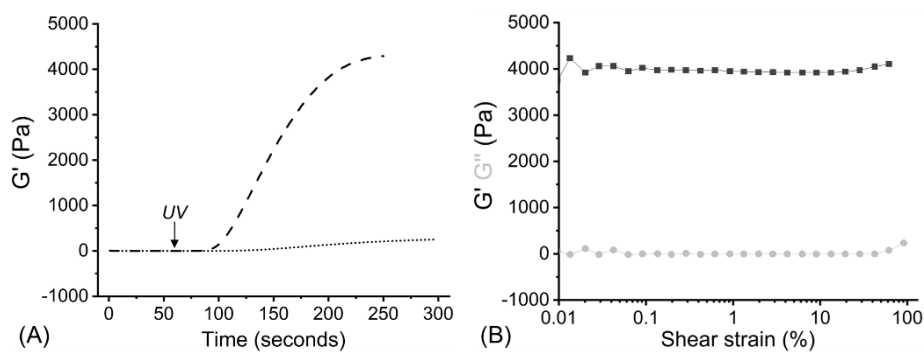

**Figure 2.** Rheological properties of thiol-ene hydrogels. (A) Oscillatory time sweep recorded during UV curing of a hydrogel-forming mixture (3.5 wt.% PEG8NB; 1 wt.% RTC-2IT; [NB]·[SH]$^{-1}$: 5.25) supplemented with either 0.1% w/v (−··) or 0.5% w/v (− −) of LAP. $G'$: storage modulus. (B) Amplitude sweep of a UV-cured thiol-ene hydrogel (0.5 % w/v LAP; 4 wt.% PEG8NB; 1 wt.% RTC-2IT; ([NB]·[SH]$^{-1}$: 6): $G'$ (−■−), loss modulus ($G''$, −●−).

The increase in $G'$ recorded following UV light activation supports the occurrence of thiol-ene reaction and the consequent synthesis of a photo-induced crosslinked network on-demand, given the incapability of the norbornene functionalities of PEG8NB to homo-polymerize. Comparable storage modulus ($G'$ ~4000 kPa) was also observed when the UV-cured hydrogels generated from thiol-ene mixture containing high LAP (0.5 % w/v) concentration was tested in oscillatory amplitude sweep (Figure 2b). Significantly lower loss modulus ($G''$) compared to $G'$ was recorded across the whole range of strain, further confirming the presence of a covalently crosslinked thiol-ene network at the molecular scale and the predominantly elastic behavior of the corresponding hydrogels. The gelation profiles of these thiol-ene hydrogels demonstrated higher storage modulus and faster gelation compared to the ones measured in previous collagen-based hydrogels in the presence or absence of a PEG crosslinking macromer,[23] supporting the validity of the presented approach towards on-demand (UV-induced) collagen network formation, irrespective of environmental factors, e.g. radical-scavenging molecular oxygen.

Following characterization of the thiol-ene gelation kinetics, the gel content, microstructure and compressive modulus of the UV-cured hydrogels crosslinked with either PEG8NB or PEG4NB were measured to assess the impact of the network architecture at the molecular, micro- and macroscale, respectively (**Figure 3**). Surprisingly, an excess of crosslinker concentration ([PEG]> 3.5 wt.%; [NB]·[SH]$^{-1}$ > 2.5) in the thiol-ene mixture was required to accomplish covalent networks with increased gel content ($G$= 58±7–93±3 wt.%) (Figure 3a), and complete solution gelation. In light of its bio-orthogonality and oxygen-insensitivity, the thiol-ene reaction is known to generate the highest yield when equimolar ratios between the sulfhydryl and vinyl groups are selected.[16] Given the non-linear configuration of both reacting macromolecules, the need for increased concentration of norbornene functionalities in the thiol-ene reacting mixture is attributed to steric effects originating from (i) the relatively small number of sulfhydryl groups (~2×10$^{-6}$ moles per g) on 2IT-reacted RTC; (ii) the large size of PEG (*Mw*: 10000-20000 g·mol$^{-1}$); and (iii) the complex architecture of both the triple helix collagen molecule and the multi-arm PEG. Overall, tuning the gel content, and, therefore, crosslink density, of thiol-ene hydrogels by simply varying the concentration of the PEG-based crosslinker opens possibilities aiming to adjust the mechanical properties in the range of skeletal muscle native stiffness, and to accomplish substrate-induced myogenesis *in vitro* and *in vivo*.[1,9,10]

Other than the variations in PEG concentration, the effect of the multi-arm architecture of PEG was also investigated. Higher gel contents were measured in thiol-ene networks crosslinked with PEG8NB ($G$= 85±4–93±3 wt.%) compared to PEG4NB ($G$= 58±7–73±3 wt.%, Figure 3a). These results agree with the increased molar content of norbornene functionalities in the reacting mixture containing PEG8NB compared to the one supplemented with PEG4NB, in line with previous reports on collagen hydrogels crosslinked with multi-arm PEGs.[23] The higher number of norbornene functionalities in PEG8NB was also effective in generating water-swollen hydrogel networks with higher, although insignificantly different, compressive modulus, compared to the use of PEG4NB (Figure 3b). This was expected in light of the direct impact of the gel content, as an indirect measure of the crosslink density, on the mechanical properties, also given that comparable porous microstructure was observed in samples crosslinked with either PEG8NB or PEG4NB (Figure 3c,d). The porosity of hydrogels is key in controlling cell response *in vitro* and *in vivo*, e.g. aiming to direct stem cell differentiation and guided bone regeneration,[6,26] and can be affected by the crosslink density, degradability, and freeze-drying process.[27]

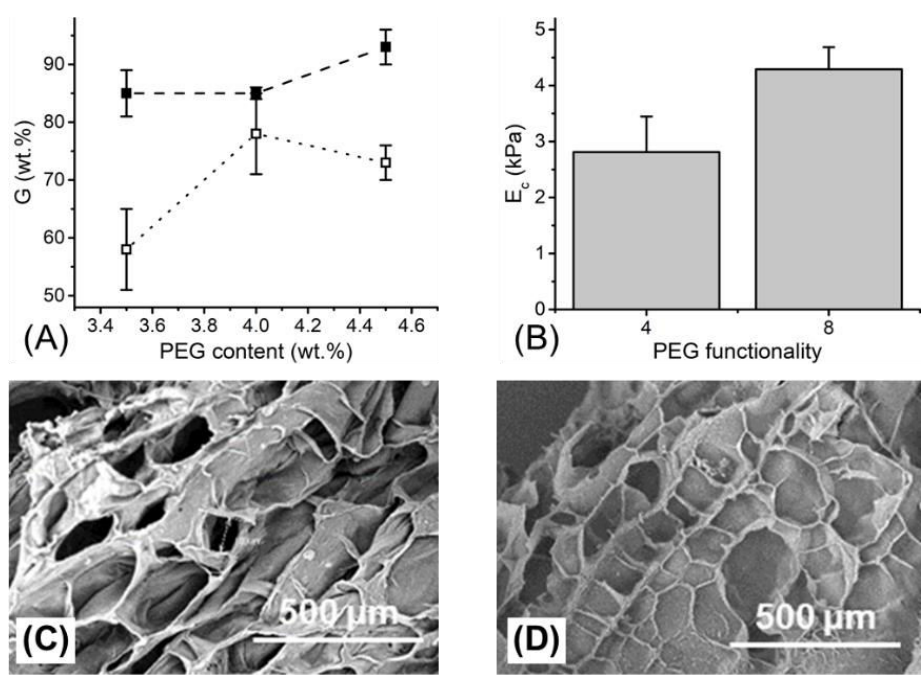

**Figure 3.** Effect of thiol-ene formulation on hydrogel properties. (A) Gel content (*G*, n=4) of photo-click hydrogels generated with varied content of either PEG4NB (f= 4, ··□··) or PEG8NB (f= 8, –■–). Lines are guidelines for the eye. (B) Compression modulus ($E_c$, n=4) of photo-click hydrogel crosslinked with either PEG4NB (3.5 wt.% PEG) or PEG8NB (3 wt.% PEG). (C-D) SEM images of the dehydrated samples reported in (B) and prepared with either PEG4NB (C) or PEG8NB (D). Fixed concentrations of thiolated collagen (1 wt.% RTC-2IT) and photo-initiator (0.5 % w/v LAP) were used.

The fact that the mechanical properties of these hydrogels can be varied independently of the porous microstructure is appealing to ensure comparable material surface for cell seeding and cell diffusion. Collectively, these results indicate that the concentration of PEG, i.e. PEG8NB, rather than the multi-arm PEG architecture, is a reliable experimental parameter aiming to accomplish thiol-ene networks with customizable crosslink density and controlled structure-property relations. Given their decreased gel content and decreased compressive modulus, samples crosslinked with PEG4NB were therefore not considered further, and further attention focused on the thiol-ene reaction with varying content of PEG8NB.

## 2.2 Macroscopic properties and degradability

The thiol-ene hydrogels crosslinked with a wider range of PEG8NB concentration and fixed amount of collagen thiol and photo-initiator (1 wt.% RTC-2IT, 0.5 % w/v LAP) were successfully tested in compression mode in the wet state (**Figure 4**a,b). Samples were coded as "CollPEGX", whereby "X" identifies the weight percent concentration of PEG8NB in the hydrogel-forming thiol-ene mixture (Figure 4c). An increase in the concentration of the PEG macromer in the thiol-ene mixture directly impacted on the compression strength and compression modulus of resulting hydrogels.

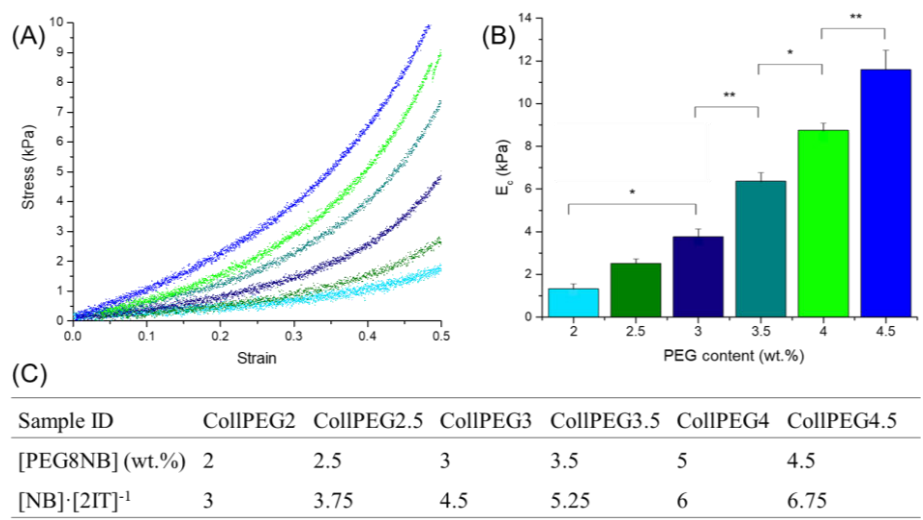

| Sample ID | CollPEG2 | CollPEG2.5 | CollPEG3 | CollPEG3.5 | CollPEG4 | CollPEG4.5 |
|---|---|---|---|---|---|---|
| [PEG8NB] (wt.%) | 2 | 2.5 | 3 | 3.5 | 5 | 4.5 |
| [NB]·[2IT]$^{-1}$ | 3 | 3.75 | 4.5 | 5.25 | 6 | 6.75 |

**Figure 4.** Compression properties of PEG8NB-based thiol-ene hydrogels in the wet state. (A) Representative stress-compression curves and (B) compression modulus ($E_c$, n=7) of samples CollPEG2 (—), CollPEG2.5 (—), CollPEG3 (—), CollPEG3.5 (—), CollPEG4 (—), and CollPEG4.5 (—). $^*p$ <0.05, $^{**}p$ <0.01. (C) Nomenclature and composition of thiol-ene hydrogels prepared with varying content of PEG8NB (f= 8) and fixed concentrations of photo-initiator (0.5 % w/v LAP) and thiolated collagen (1 wt.% RTC-2IT).

Significantly different compression moduli were recorded in hydrogels CollPEG3, CollPEG3.5, CollPEG4 and CollPEG4.5 ($E_c$= 3.8±0.4–11.6±0.9 kPa), whilst low mechanical properties ($E_c$ <3 kPa) were exhibited by samples CollPEG2 and CollPEG2.5 (Figure 4c), likely due to the relatively low concentration of norbornene functionalities in the reacting mixture. The aforementioned results, therefore, agree with previous trends in gel content measured in the corresponding UV-cured networks and confirm the formation of covalently crosslinked thiol-ene hydrogels in light of the impossibility of PEG8NB to homo-polymerize. All samples revealed a *J*-shaped stress-strain curve during compression, reflecting the strain-hardening behaviour of collagen-based materials.[6,28,29,30] At the same time, no mechanical break was observed following hydrogel compression up to 50%, unlike the case of previously reported UV-cured methacrylated collagen networks.[28b,29] This observation agrees with the selectivity of the thiol-ene crosslinking reaction, and the increased homogeneity (and compressibility) of respective network architecture.[31] Given that the compression tests were carried out in the water-swollen samples, swelling studies were therefore carried out to understand the behavior of this material in a near-physiologic environment.

**Figure 5** displays the gravimetric swelling profiles recorded with the different samples in distilled water.

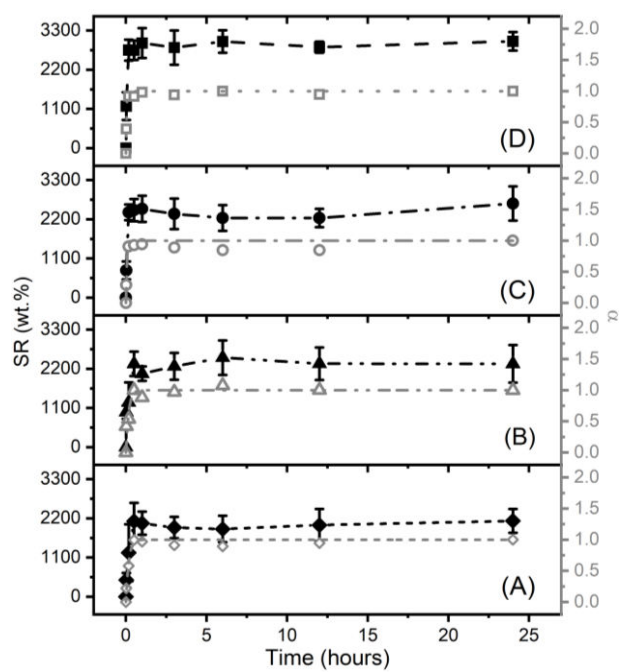

**Figure 5.** Swelling profiles of thiol-ene hydrogels in distilled water at room temperature. (A) CollPEG2.5; (B) CollPEG3; (C) CollPEG 3.5; (D) CollPEG4. Black scatters and the left-hand side y axis indicate the value of the swelling ratio (*SR*, n=4); black curves are guidelines for the eye. Grey scatters and the right-hand side y axis describe the normalised *SR* data with respect to equilibrium values ($\alpha = SR_t/SR_{eq}$). Grey curves are the first order fit of $\alpha$.

A significant uptake of medium was measured following the first hour of dry network incubation, so that an averaged swelling ratio (*SR*) higher than 2000 wt.% (*SR*= 2063±326–2949±419 wt.%) was obtained with no significant variations at later incubation time points. An increase in the PEG8NB concentration in the thiol-ene reacting mixture was found to directly affect the *SR* of respective hydrogels, so that samples CollPEG4 (*SR*= 2999±262 wt.%) and CollPEG2.5 (*SR*= 2129±337 wt.%) revealed the highest and lowest weight increase after 24 hours, respectively, although with an insignificant difference (*p* =0.0877). Marginal variations in swelling values (*SR* =3010±127 wt.%, *p* ≥0.05) were also measured in sample CollPEG4.5 following 24-hour incubation, so that hydrogels prepared with increased PEG8NB content ([NB]·[2IT]$^{-1}$ >4) were not investigated further. Surprisingly, these swelling trends appeared to be directly related to changes in *G* (Figure 3a), whereby samples prepared with higher concentrations of PEG8NB exhibited an increased crosslink density and yet increased swelling degrees. Considering the orthogonality of the presented thiol-ene reaction, these results therefore suggest that the presence of the PEG crosslinker mediates the diffusion and retention of water in the corresponding covalent networks, yielding higher values of *SR* compared to the case of PEG-free UV-cured RTC-based hydrogels.[28,29] To elucidate this point, the swelling kinetics of the thiol-ene networks were analyzed.

The swelling of hydrogels typically includes (i) the diffusion of water molecules, (ii) the hydration and relaxation of polymer chains, and (iii) the expansion of the polymer network.[32] To assess whether water diffusion or chain relaxation played a major role, the swelling profiles of the thiol-ene hydrogels were fitted with the Peppas model and first-order kinetics (**Table 1**, Figure 5). The Peppas model revealed good description of the swelling process ($R^2$> 0.83, **Figure S1**), whereby low values of swelling coefficient (n <0.5) were obtained, indicating that the water penetration rate in the network is slower than the polymer chain relaxation rate.

**Table 1.** Kinetic constants for hydrogel swelling following data fitting according to Peppas model ($\alpha = k \cdot t^n$) and first-order kinetics ($\alpha = 1 - e^{-k \cdot t}$).

| Sample ID | Peppas model | | | First-order kinetics | |
|---|---|---|---|---|---|
| | k | n | $R^2$ | k | $R^2$ |
| CollPEG2.5 | 1.3636 | 0.4626 | 0.9984 | 5.9432 | 0.9613 |
| CollPEG3 | 1.1538 | 0.3029 | 0.8438 | 6.0097 | 0.8727 |
| CollPEG3.5 | 1.2265 | 0.2799 | 0.8348 | 17.9999 | 0.9270 |
| CollPEG4 | 1.1542 | 0.2205 | 0.8316 | 28.9872 | 0.9803 |

This situation suggests a Fickian diffusion mechanism, which confirms that the high hydrophilicity of both type I collagen and PEG[33] promotes increased binding of water molecules with, and therefore increased water uptake, in the covalent network. The diffusion-driven swelling mechanism was also supported by data fitting via first-order kinetics (Figure 5), whereby close data representation was confirmed across the whole range of data points ($R^2$ >0.87, Table 1).

As well as the swelling properties, the hydrolytic and enzymatic stability of the thiol-ene hydrogels were assessed over a 7-day incubation period through gravimetric analysis and compression testing (**Figure 6**). Samples CollPEG3.5, CollPEG4 and CollPEG4.5 were selected due to their relatively high gel content ($G$ >85 wt.%, Figure 3a), in contrast to the incomplete gelation observed with samples CollPEG2.5 and CollPEG3, consequent to the lower molar content of PEG8NB in respected thiol-ene mixtures.

All samples revealed minimal mass loss and insignificant variations in wet state compression modulus in hydrolytic conditions, while up to 50% mass loss was observed in sample CollPEG3.5 following incubation in a collagenase-supplemented medium (Figure 6a), in agreement with the enzymatic peptide scission of collagen molecules in the covalent network.

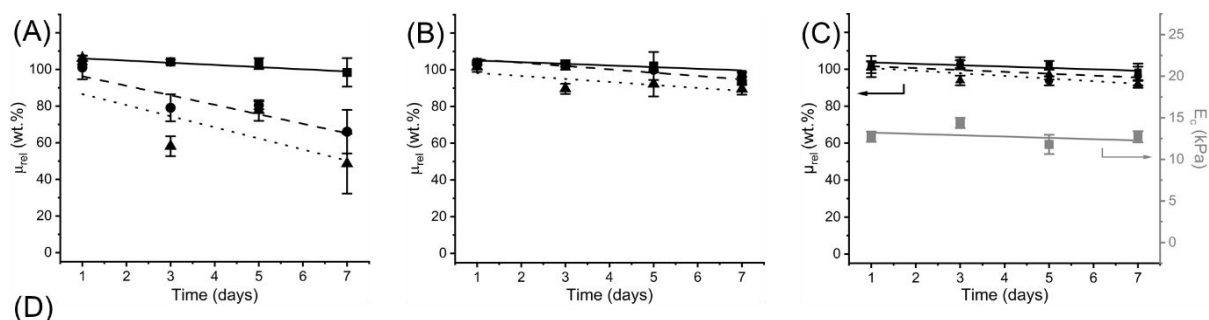

| Sample ID | $\mu_{rel}(E_c) = a + b \cdot t$ (HEPES-BBS) | | | $\mu_{rel}$ (0.2 mg·ml$^{-1}$ CLG) | | | $\mu_{rel}$ (2 mg·ml$^{-1}$ CLG) | | |
|---|---|---|---|---|---|---|---|---|---|
| | a | b | $R^2$ | a | b | $R^2$ | a | b | $R^2$ |
| CollPEG3.5 | 107.20 | -1.19 | 0.77 | 101.46 | -5.19 | 0.89 | 92.62 | -6.06 | 0.53 |
| CollPEG4 | 105.97 | -0.89 | 0.67 | 107.52 | 1.81 | 0.90 | 99.93 | 1.63 | 0.58 |
| CollPEG4.5 | 104.42 (13.37) | 0.74 (-0.16) | 0.62 (0.12) | 102.08 | 1.42 | 0.71 | 102.61 | -1.01 | 0.57 |

**Figure 6.** Degradation of photo-click hydrogels *in vitro* (37 °C, pH 7.4). (A): Coll-PEG3.5; (B): Coll-PEG4; (C): Coll-PEG4.5. (■, —) and (■, —): remaining mass ($\mu_{rel}$) and wet state compression modulus ($E_c$) in HEPES-BBS, respectively. (●, – –) and (▲, ⋯): $\mu_{rel}$ in HEPES-BBS supplemented with either 0.2 mg·ml$^{-1}$ or 2 mg·ml$^{-1}$ collagenase (CLG), respectively. Lines indicate the linear fitting of single data. (D): linear regression of $\mu_{rel}$ and $E_c$ (in bracket).

Linear fitting proved to accurately describe the degradation process in both hydrolytic and enzymatic environments (Figure 6d), indicating a surface rather than bulk erosion mechanism, with the former being observed with other type I collagen-based hydrogels.[6,29] These results therefore proved to be inversely related to previous trends in gel content (Figure 3), as additional evidence of the presence of a covalent network in the thiol-ene hydrogels. Consequently, it was possible to leverage variations in gel content and crosslink density to accomplish hydrogels with controlled degradation in biological conditions.

Considering the skeletal muscle tissue regeneration process *in vivo*, the erosion degradation mechanism revealed by the presented hydrogels is appealing, aiming to minimise abrupt changes in macroscopic properties and ensure a gradual transmission of mechanical loads from the degrading material to the skeletal muscle neo-tissue. Furthermore, given that the muscle regeneration process can take up to eight weeks,[34] sample CollPEG4.5 appeared to display a suitable degradation profile, given its high mass retention *in vitro* ($\mu_{rel}$ >90 wt.%) and the linear decrease in compression modulus observed over the 7-day period.

## 2.3 Cytotoxicity evaluation *in vitro*

The elasticity of the ECM is an intrinsic physical characteristic that has been shown to have a profound effect on the spreading, morphology and differentiation of mesenchymal stem cells (MSCs),[35] as well as in tissue pathology.[20] MSCs show lineage-specific differentiation when cultured on substrates matching the stiffness of native tissues. Soft tissue-like matrices ($E_c$ ~2 kPa) tend to induce MSC differentiation into neural cells, whereas stiffer matrices ($E_c$ ~10 kPa) induce myocyte formation.[36] At increased content of PEG8NB ([NB]·[2IT]$^{-1}$ ≥3.5), the thiol-ene hydrogels revealed a range of compressive modulus ($E_c$ =6.4±0.4–11.6±0.0 kPa, Figure 4) matching the skeletal muscle native stiffness ($E_c$ ~8-11 kPa),[10] and that is why they were selected as stem cell niches to study the myogenic differentiation *in vitro*.

Our strategy to test the hydrogel-induced myogenesis *in vitro* relied on the use of C2C12 murine myoblasts, due to their extensive use in muscle biology, and their rapid proliferation and differentiation in cell culture systems.[37] An extract and a direct cytotoxicity test were initially performed, whereby the number of cells was quantified via ATP assay (**Figure S2**), to assess the tolerability of these cells to the thiol-ene hydrogels. Although both type I collagen and PEG are approved by the FDA, these cytotoxicity tests were considered key, in light of the relatively high, potentially toxic concentration of the photo-initiator used in the UV curing step.[23] ATP assay following 2-day cell culture on the hydrogel extracts (**Figure 7**a) revealed no significant difference between the number of cells grown on the hydrogels compared to the cell culture medium (positive control), whilst no cell was observed in the negative control in the same conditions. These results therefore speak against the leaching of any soluble, cytotoxic compounds from the above hydrogel to C2C12 cells, an observation that is supported by the high gel content ($G= 85\pm4–93\pm3$ wt.%) measured on the same samples (Figure 3a).

Further to indirect testing using extracts, cell culture seeding was performed to examine the direct cytotoxicity of the hydrogels on cells *in vitro* over a 5-day cell culture period. Initially, live /dead cell imaging was carried out to assess cell attachment (following 3 hours of cell seeding) and spreading (following 24 hours of cell seeding). Comparable cell attachment was observed across the three hydrogels, which proved to be significantly lower compared to the one found on the tissue culture plastic (Figure 7b).

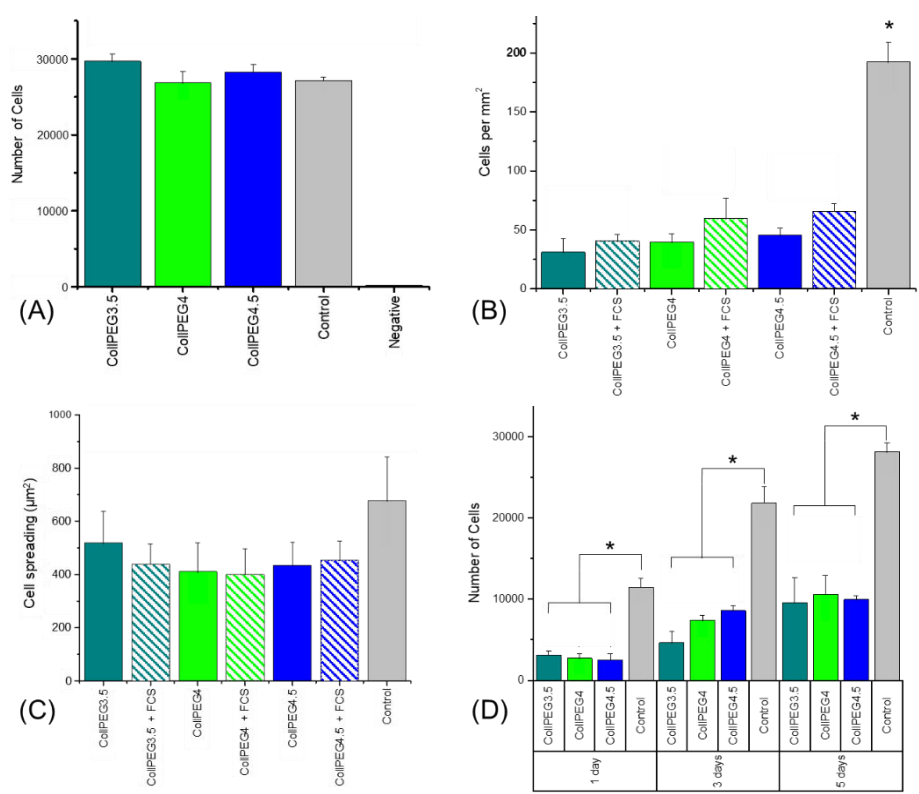

**Figure 7.** Cytotoxicity tests of thiol-ene hydrogels with C2C12 murine myoblasts. (A): ATP assay (n=5) carried out following 2-day cell culture on hydrogel extracts, DMEM (positive control) and DMSO-supplemented DMEM (negative control). (B): Cell attachment (n=3) following 3-hour cell culture onto FCS-incubated and pristine hydrogels (*$p$ <0.05). (C): Cell spreading (n=3) following 24-hour cell culture onto FCS-incubated and pristine hydrogels. (D): ATP assay (n=5) carried out during a 5-day cell culture on the photo-click hydrogels and tissue culture plastic control (*$p$ <0.05).

However, cell spreading recorded at a longer time point revealed insignificantly differences among all sample groups and the control (Figure 7c). Here, hydrogel incubation in media supplemented with FCS-associated cell-binding proteins[38] did not appear to induce any detectable effect on either cell attachment or cell spreading, indicating that the type I collagen in the covalent network is the sole factor in controlling cell response. Collectively, these observations, therefore, suggest that hydrogels can effectively support cell spreading despite the initial decreased cell attachment compared to the tissue culture plastic control group. The preferential early cell attachment to the control group was as expected, since tissue culture plastics undergo plasma treatment during manufacture specifically to ensure such cell attachment.

Attention subsequently moved to assessing C2C12 proliferation (via ATP assay) on the hydrogel at day 1, 3 and 5 of cell culture (Figure 7d). At each time point, the number of cells measured on the hydrogels was lower than the one measured on the tissue culture plastic control. No cell was detected in the media alone, blank, native hydrogel, and the negative control group (**Figure S3**), confirming that the observed cell responses were attributed solely to the hydrogel characteristics. The number of cells measured on each hydrogel was found to increase at longer times of cell culture, so that significant differences were measured between day 1 and day 5 in samples of CollPEG4 ($p$ =0.0319) and CollPEG4.5 ($p$ =0.0460). These observations therefore suggest that the hydrogels support the proliferation of C2C12 cells, while the higher number of cells on the tissue culture control compared to the collagen material is attributed to the increased cell adhesion (Figure 7b), rather than to a toxic effect triggered by the hydrogel.

Ultimately, there was no statistically significant difference in the number of cells measured among the three hydrogels, although sample CollPEG3.5 tended to show lower values. This observation indicates that an increased molar content of the synthetic PEG phase, i.e. free of cell-binding sequences, in the hydrogel did not induce any significant effect on cell viability, so that it was possible to adjust the mechanical and swelling properties without affecting the cell response.

**2.4 Cell maturation and differentiation**

Having confirmed the C2C12 tolerability of selected thiol-ene hydrogels, the next step was to assess hydrogel-induced cell maturation/ differentiation, which is the process through which myoblasts switch from a proliferating state to a fusing state to form myotubes. Cells were grown up to 70% confluency and then transferred to a low serum media (**Figure 8**), whereby a first phase of actin remodelling followed by an increase in fusion index (after 3 days) can be observed.[39] This change in the cell morphology, including nuclear fusion, was visualised in the fluorescent images captured from day 0 to day 3 and 7 of cell culture.

At day 0 (70% confluence), similar cell morphology was observed on the three hydrogels (Figure 8a,d,g) and tissue culture plastic control prior to change in culture medium (Figure 8j), an observation that agrees with the comparable cell spreading observed following 24-hour cell culture on the different sample groups (Figure 7c).

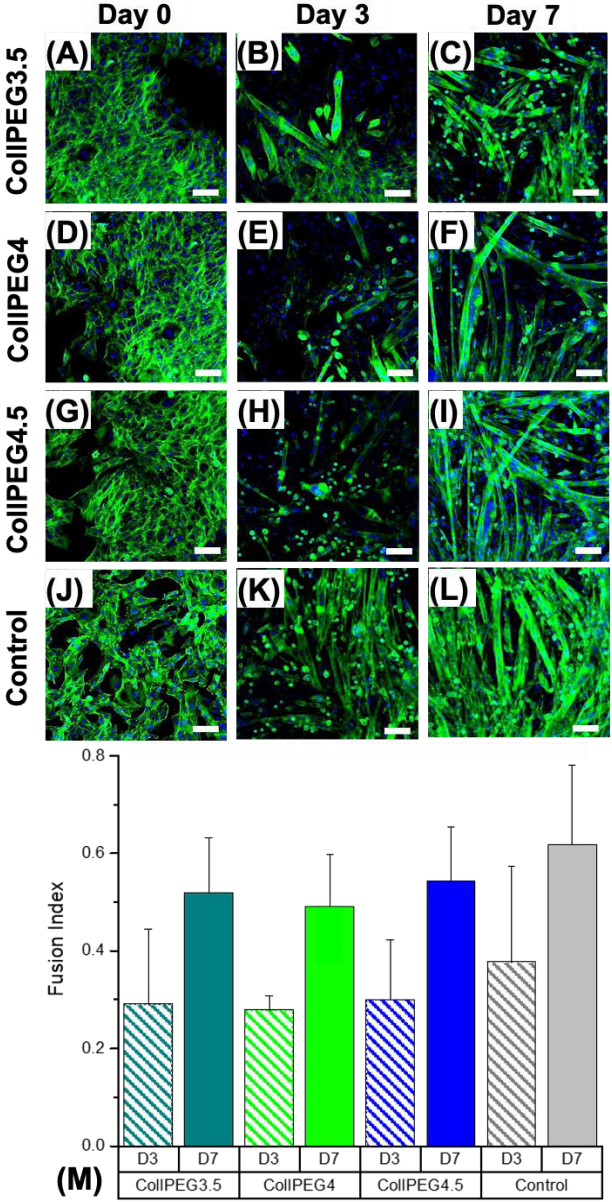

**Figure 8.** Immunofluorescent images describing the maturation and myogenic differentiation of C2C12 cells following 7-day culture on the thiol-ene hydrogels or tissue culture plastic control in the presence of low serum medium (DMEM, 1% PenStrep, 1% FCS, 1% ITS). (A-C): CollPEG3.5; (D-F): CollPEG4; (G-I): CollPEG4.5; (J-L): tissue culture plastic control. Green: F-actin, blue: nuclei. Scale bar ≈ 100 μm. (M): Fusion index calculated on day 3 (D3) and day 7 (D7) of cell culture on either the hydrogels or the tissue culture plastic control.

At the same time point, the myoblasts appeared less spread on the hydrogels compared to the tissue culture plastic control, an observation that reflects previously discussed trends in cell attachment (Figure 7b) and proliferation (Figure 7d). Cells were found to fuse together at increased tissue culture time points, so that myotube formation was detectable from day 3 in a random alignment. At day 3, the viability of cells on the collagen samples (Figure 8b,e,h) appeared to be lower than that of the control group (Figure 8k). This result is attributed to the cell washing prior to transfer to the low serum cell culture medium, potentially inducing cell removal and, therefore, resulting in a decreased number of cells being cultured. This explanation is supported by the immunofluorescent images captured on day 5 (Figure 8c,f,i), whereby all three hydrogels showed myotube formation and increased cell viability compared to day 3 (Figure 8b,e,h), confirming previously observed cell proliferation trends (Figure 7d). It was noted that the myoblasts appeared less spread on the hydrogels at day 0 (Figure 8a,d,g) compared to the tissue culture plastic control (Figure 8j), an observation that agrees with previously discussed differences in cell attachment (Figure 7b). Indeed, once the cells began to fuse together, this observation was no longer evident and myotube formation was detectable from day 3 in a random alignment.

No statistical difference in fusion index (*FI*) was observed between either the three hydrogels (*FI* =0.28±0.03–0.54±0.11, *p* =0.83–1.00) or the tissue culture plastic control at day 3 (*FI* =0.38±0.19, *p* =0.98–1.00) or day 7 (*FI* =0.62±0.15. *p* =0.61–0.99) of maturation/ differentiation (Figure 8m), although higher values were recorded in the control group. Comparable measurements of *FI* were observed on satellite cells following 7-day culture on a mechanically comparable hydrogel consisting of a blend of chitosan and type I collagen (tensile modulus =12±1 kPa, *FI* =0.90±0.06), on the one hand, and tissue culture plastic (*FI* =0.60±0.06), on the other hand.[40] The lower, although insignificantly different (*p* =0.1363), value of *FI* recorded with sample CollPEG4.5 ($E_c$ =11.6±0.9 kPa, *FI* = 0.54±0.11) is likely attributed to differences in cell type and material composition.

Immunofluorescent labelling confirmed that there was no positive staining for myosin at day 3 in all three hydrogels and the control group (**Figure 9**), in contrast to previous observations obtained via f-actin staining (Figure 8). However, cells cultured on the three hydrogels were stained positive for myosin at day 7 (Figure 9b,d,f), similarly to the case of the control group (Figure 9h). The different observations revealed by the two aforementioned types of staining are in agreement with previous reports[41] and reflect the kinetics of myogenic maturation /differentiation, whereby the initial actin filament remodeling is a precursor to myotube formation.

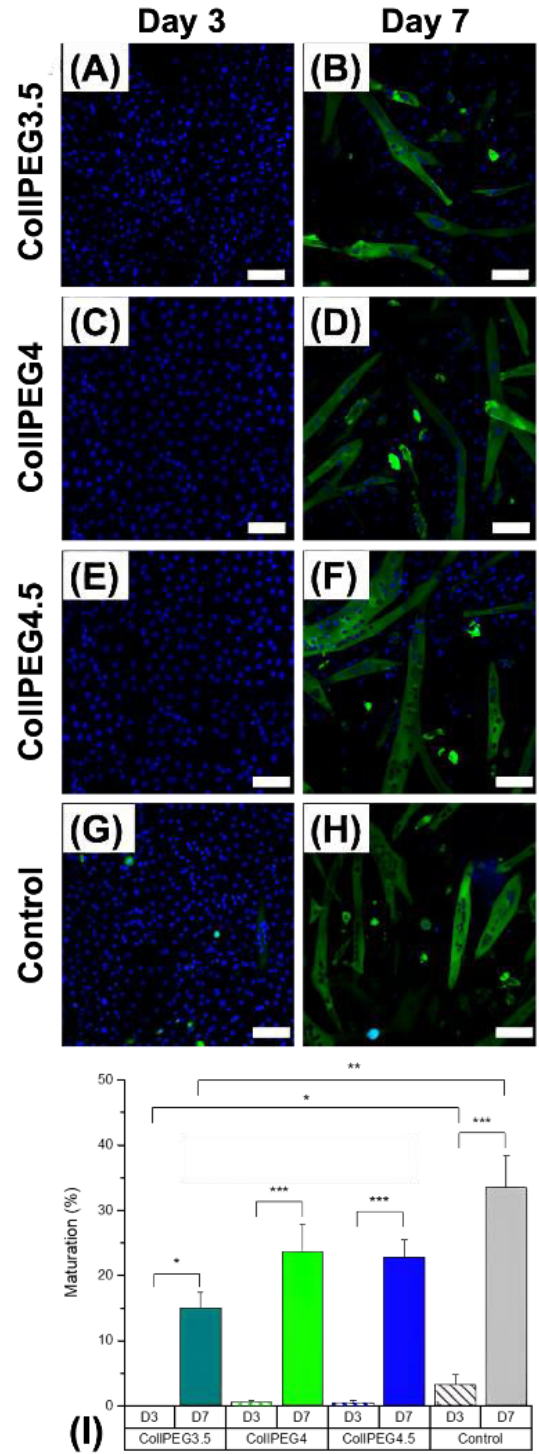

**Figure 9.** Immunofluorescent images indicating myosin expression at days 3 and 7 of C2C12 cell culture on the thiol-ene hydrogels and tissue culture plastic control. Green: myosin heavy chain; blue: cell nucleus. (A-B): CollPEG3.5, (C-D): CollPEG4; (E-F): CollPEG4.5; (G-H): tissue culture plastic control. Scale bars ≈ 100 μm. (I): Myoblast maturation observed on day 3 (D3) and day 7 (D7) of cell culture ($^{*} p \leq 0.05$, $^{**} p \leq 0.01$, $^{***} p \leq 0.005$).

Cell maturation was quantified for the three hydrogels and the tissue culture plastic control at day 3 and 7 (Figure 9i). As expected, increased values were recorded at later time points, whereby the highest maturation was measured on the tissue culture plastic throughout the analysis in line with previous observations on cell attachment, spreading and proliferation (Figure 7b-d). At day 3, the control revealed significantly increased myosin expression compared to sample CollPEG3.5 ($p \leq 0.05$), and no statistical significance was found between the three hydrogels ($p =0.98–1$). By day 7, all the groups had promoted significantly higher C2C12 cell maturation compared to the earlier time points ($p =0.0003–0.0196$), whereby increased myosin expression was observed on the control group compared to sample CollPEG3.5 ($p =0.0034$). Although selected hydrogels exhibited significantly different mechanical properties (Figure 4), this variation was not resolved at the cell maturation level. Indeed, similar maturation ($p= 0.37–0.48$) was found between the three mechanically different hydrogels, indicating comparable proliferation of cells across the three hydrogels. These results may therefore suggest that the variation in compression modulus across the three hydrogels is too narrow to enable a detectable effect at the cellular scale.

**2.5 Subcutaneous study in rats**

Medical devices initiate a host response once implanted in *vivo*, which spans from hours to days depending on the interaction of the biomaterial with the immune system. The host response is not known *a priori*, and can only be assessed via appropriate testing *in vivo*. Given the novelty of the thiol-ene hydrogels developed in this study, it was of interest to investigate the biocompatibility *in vivo*, especially following successful testing *in vitro*. A subcutaneous model in rats was therefore used to analyse the local immune response to the hydrogel. Sample CollPEG4.5 was selected for the implantation *in vivo* since its mechanical properties proved to mimic native skeletal muscle stiffness (Figure 4). This hydrogel also showed the lowest degradation *in vitro* (Figure 6), which is appealing considering that the muscle regeneration process can take up to eight weeks.[34] Mucograft® (Geistlich) was chosen as the clinical gold standard collagen matrix as it is widely employed for soft tissue regeneration. Samples were collected at day 1, 4 and 7 following the subcutaneous implantation of the hydrogel, aiming to assess the different immune responses. Innate immunity is the initial response (hours) to an antigen, whereas the adaptive immunity occurs after several days and results in the activation, differentiation and expansion of lymphocytes.[42] From these events, acute inflammation develops in minutes and disappears in days, whereas chronic inflammation stems from prolonged tissue injury.

**Figure 10** reports the H&E histology images at different time points during the 7-day implantation study. After 1 day, the hydrogel proved to be surrounded by a layer of cells (Figure 10a), whilst minimal cell integration was detectable near Mucograft® (Figure 10b). While no evidence of polynuclear neutrophils was obtained, the surrounding protective matrix that was secreted around the hydrogel suggests an early innate immune response, in line with the implantation of the exogenous material.

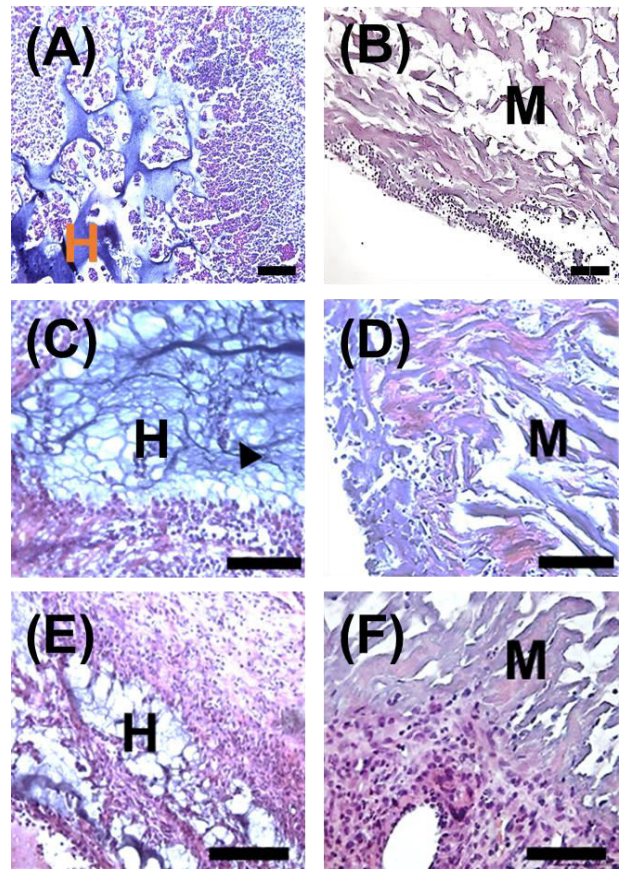

**Figure 10.** H&E stain of samples collected at day 1 (A-B), 4 (C-D) and 7 (E-F) following subcutaneous implantation in rats. H: hydrogel CollPEG4.5 (A, C, E); M: Mucograft® (B, D, F);. Scale bars ≈ 50 μm. The black arrow points to "fried egg" shaped macrophages.

After 4 days of implantation, a hydrogel-induced immune response was apparent as indicated by the presence of a dense layer of surrounding cells (Figure 10c). Cell infiltration in, and debridement of, the hydrogel was also evidenced by the presence of "fried egg" shaped macrophages (black arrow). In comparison, no tissue attachment was still observed around the sample of Mucograft® (Figure 10d), with only a small amount of infiltrated round-like nuclei rather than cytoplasm-surrounded cells. These observations are in agreement with previously-reported tissue response to Mucograft® in CD-1 mice, revealing no evidence of capsule formation around the material at day 3 post-implantation.[43]

After 7 days *in vivo*, a dense layer of cells was still detected on the surrounding of hydrogel

CollPEG4.5, although to a less extent compared to day 4 (Figure 10e). Cell infiltration through the material was detectable, together with the formation of connective tissue, as depicted by the presence of spindle-shaped fibroblasts, as well as muscle tissue. Fibroblasts appeared to proliferate at the injury site and remodel the local ECM to repair the wound, consistently with the case of an acute foreign body reaction. On the other hand, the sample of Mucograft® displayed higher integration with the surrounding tissue compared to the earlier time points, as well as the formation of blood vessels. Although no dense population of surrounding inflammatory-like cells was detected, an increased number of infiltrating cells was found compared to the earlier time points, whereby a round-like nucleus, rather than cytoplasm-surrounded, configuration was again visible, in agreement with previous study.[37]

The thiol-ene hydrogels presented in this work were designed to mimic skeletal muscle native elasticity, and, ultimately, to support tissue repair and encourage biomaterial integration. While Mucograft® presented a lower immune response and reduced inflammation, this investigation *in vivo* indicated a low dimensional variation in sample CollPEG4.5 after 7 days, in line with the previously measured high gel content (Figure 3) and degradation profile *in vitro* (Figure 6). Compared to the commercial benchmark, the hydrogel appeared to trigger an increased immune response, which was beneficial to promote angiogenesis and a dense interface with the host tissue, which could be key to support muscle regeneration at the later time points of implantation. No sign of hydrogel-induced chronic inflammation or macrophage-fused foreign body giant cells was observed during the 7-days *in vivo*, which is in agreement with either extract or contact cytotoxicity results *in vitro* (Figure 7), supporting the biocompatibility of this material in biological environments.

## 3. Conclusions

A novel platform of photo-click hydrogels was successfully developed via thiol-ene reaction between 2IT-reacted RTC and multi-arm norbornene terminated PEG with varied number of norbornene functionalities. The hydrogels proved to promptly form following UV activation, whereby variation in gelation kinetics, as well as gel content and compression modulus was demonstrated by changes in photo-initiator concentration, and changes in the concentration and molecular architecture of PEG, on the other hand. An excess in PEG concentration was required to accomplish full solution gelation and tunable macroscopic properties, likely due to steric hindrance effects between the complex thiolated collagen molecules and the multi-arm architecture of the PEG crosslinker. No cytotoxic effects were observed via either extract or contact cytotoxicity tests. Hydrogels proved to support cell spreading, proliferation, as well as myotube formation, although the narrow variation in compression moduli were not reflected in differences at the cell level. The material was found to invoke a higher immune response compared to a commercial gold standard in rats, which proved key to ensure increased cell infiltration, blood formation and cell density at the tissue interface. The presented hydrogels therefore present features that make them appealing for muscle tissue applications, although further preclinical testing is required to assess their capability in soft tissue regeneration.

## 4. Experimental Section

*Materials*: rat tails were acquired post-mortem from the University of Leeds' animal house. Lithium phenyl-2,4,6-trimethylbenzoylphosphinate (LAP) was purchased from Tokyo Chemicals Industry. 8-arm norbornene-terminated PEG (PEG8NB, f= 8, $M_w$: 20,000 g·mol$^{-1}$) was supplied by JenKem Technology USA. Acetic acid, collagenase from Clostridium histolyticum, 4-arm norbornene-terminated PEG (PEG4NB, f= 4, $M_w$: 10,000 g·mol$^{-1}$), 4-(2-hydroxyethyl)-1-piperazineethanesulfonic acid (HEPES), 4',6-diamidino-2-phenylindole (DAPI) and Horse Serum were purchased from Sigma Aldrich. Dithiothreitol (DTT) was purchased from Life Technologies. 2-iminothiolane hydrochloride (2IT), Dulbecco's modified eagle medium (DMEM), penicillin-streptomycin, and Oregon Green 488 Phalloidin were supplied by Fisher Scientific Ltd. (Loughborough, UK). Myosin 4 Antibody Alexa Fluor 488 (MF20) and Insulin-Transferrin-Selenium (ITS) were purchased from Thermo Fisher Scientific.

*Synthesis of thiolated collagen*: in-house extracted rat tail collagen (RTC)[28] was solubilized in a 17.4 mM acetic acid solution (1 wt.% RTC) via magnetic stirring at room temperature. The pH of the solution was adjusted to pH 7.4 and then 2IT ([2IT]·[Lys]$^{-1}$= 15) and DTT ([DTT]·[2IT]$^{-1}$= 1) were added. The reaction was run at room temperature for 24 hours under magnetic stirring and terminated through precipitation in ethanol (20-fold volume excess). The thiolated collagen product (RTC-2IT) was retrieved via centrifugation (10,000 rpm, 30 min, 4 °C) and air dried. Covalent coupling of 2IT was estimated via (2,4,6)-trinitrobenzenesulfonic acid (TNBS) colorimetric assay,[44] which demonstrated an averaged consumption of primary amino groups of 72 mol.%.

*Preparation of thiol-ene photo-click hydrogels:* RTC-2IT (1 wt.%) was dissolved in 10 mM phosphate buffered solution (PBS) supplemented with LAP (0.1-0.5% w/v). Either PEG4NB or PEG8NB (2-4 % w/v PEG) was added to the collagen solution and exposed to UV light (Spectroline, 365 nm). Formed hydrogels were washed in distilled water and dehydrated in an ascending series of distilled water-ethanol mixture (0-100 vol.% EtOH).

*Characterisation*: dry samples of known weight ($m_0$: 2-5 mg) were individually placed in 1 mL of distilled water at 25 °C under mild shaking. At the specific time point, swollen samples were collected, paper blotted and weighed ($m_s$). The swelling ratio (*SR*) was calculated according to Equation 1:

$$SR = \frac{m_s - m_0}{m_0} \times 100 \tag{1}$$

Swelling data were fitted according to Peppas and the first-order kinetics models, as described by Equation 2 and Equation 3, respectively:

$$\alpha = k \cdot t^n \tag{2}$$

$$\alpha = 1 - e^{k \cdot t} \tag{3}$$

where $\alpha$ is the ratio of the *SR* at selected time points with respect to the *SR* at equilibrium; *n* is the swelling exponent describing the mechanism of solvent transport into the hydrogel; and *k* is the diffusion coefficient of water into the hydrogel.

In addition to the *SR*, the gel content (*G*) was determined to quantify the overall fraction of the insoluble, covalently crosslinked network. Dried samples of known mass ($m_0$) were incubated in 10 mM HCl solution for 24 h, aiming to remove any soluble phase dispersed in the network. Samples were then rinsed in distilled water, air-dried and weighed ($m_e$). The *G* was calculated according to Equation 4:

$$G = \frac{m_e}{m_0} \times 100 \tag{4}$$

Four replicates of each sample group were employed for the quantification of the *SR* and *G* and the results were expressed as mean ± standard error.

The microstructure of formed hydrogels was investigated with scanning electron microscopy (SEM) using a cool stage (JEOL SM-35). SEM images were captured via backscattered electron detection at 5 kV and 12-13 mm working distance.

*Rheology*: UV-curing time sweeps were performed at 25 °C to assess the gelation kinetics of LAP-supplemented hydrogel-forming mixtures of RTC-2IT and PEG8NB. A UV light-equipped rheometer was employed (Anton Paar MCR 301) with a cone plate (CP50-2, ø: 49.97 mm, cone angle: 1.996°). The UV light (365 nm, 4450 µW·cm$^{-2}$) was initiated 60 seconds after the start of the measurements and variations in shear moduli were recorded. Amplitude sweeps (100 rad·s$^{-1}$, 1 N) were performed at 25 °C on the resulting photo-click hydrogels using a parallel plate (ø: 15 mm).

*Compression tests*: water-equilibrated hydrogel discs (Ø: 12 mm) were compressed (Instron ElectroPuls E3000) at room temperature with a compression rate of 3 mm·min$^{-1}$ and a 250 N load cell. Stress-strain curves were recorded, and the compression modulus was quantified via linear fitting at 0-20% compression. Up to seven replicates were tested for each sample group, and data expressed as mean ± standard error.

*Degradation tests*: dry hydrogels of known weight ($m_0$) were placed in a HEPES-buffered saline solution (HEPES-BSS, pH 7.4) supplemented with 0, 0.2 or 2 mg·ml$^{-1}$ collagenase and incubated at 37 °C for up to 7 days. At a specific time point, samples were removed, rinsed with distilled water and dried prior to weight measurement ($m_d$). The collagenase-supplemented medium was replaced every two days to minimise risks of enzyme activity loss. The relative mass at each specific time point (*t*) was determined according to Equation 5:

$$\mu_{rel}(t) = \frac{m_d(t)}{m_o} \times 100 \qquad (5)$$

Linear regression of the data was performed to gain insight into the degradation mechanism, either via surface or bulk erosion.

In addition to gravimetric analysis, the wet samples collected at each degradation time point were also analysed under compression, as described above.

Three replicates were used for each time point, and data expressed as mean ± standard error.

*Cytotoxicity tests*: Both extract and direct cotact cytotoxicity tests (ISO 1993-5) were carried out with C2C12 mouse C3H muscle myoblasts (Sigma Aldrich). Cells were incubated (37 °C, 5% $CO_2$) in DMEM supplemented with foetal calf serum (FCS, 10%) and Penicillin Streptomycin (PenStrep, 1%), to reach 50% confluence in a 96-well plate. The dry samples of CollPEG3.5, CollPEG4 and CollPEG4.5 were gamma sterilised (29-38 kGy), and a fraction of these sterilised samples treated in FCS-supplemented DMEM (50 vol.%) overnight (samples coded as CollPEG3.5 + FCS, CollPEG4 + FCS and CollPEG4.5 + FCS).

For extract cytotoxicity tests, sample extracts were prepared by incubating the dry material (1 mg) in cell culture media (1 mL) for 72 hours. The sample extract was recovered by centrifugation and applied to 50% confluent C2C12 cells cultured on a polystyrene 96-well plate. A mixture (50 vol.%) of dimethyl sulfoxide (DMSO) in DMEM was used as the negative control, whilst cell culture media with 10% FCS was used as the positive control. After 48-hour culture, cell viability was assessed using an ATPLite luminescence assay (PerkinElmer), according to manufacturer instructions. Five replicates were used for each sample group, and the number of cells expressed as mean ± standard error.

For direct cytotoxicity tests, the photo-click hydrogels (100 μL) were prepared in a 96-well plate, air-dried and gamma-sterilised. The sterilised materials were swollen in cell culture media in a 96-well plate, and C2C12 mouse myoblasts (1×10$^4$ cells) were applied to each hydrogel. A mixture of DMSO (50 vol.%) in DMEM was used as the negative control, and cells cultured on tissue culture plastic were selected as the positive control. After 1, 3 and 5 days of culture, the hydrogels were transferred to a fresh well plate, and cell viability was assessed via ATPLite luminescence assay. Five repeats were used for each sample group, and the number of cells expressed as mean ± standard error.

*Cell attachment, spreading and proliferation:* To quantify cell attachment, viable C2C12 cells cultured in a T175 flask were labeled with calcein AM according to manufacturer instructions (Life Technologies). The cells were trypsinised and re-suspended in cell culture media. Cells were subsequently seeded onto either FCS-treated or untreated gamma-sterilised hydrogels (n=3; $1\times10^5$ cells·mL$^{-1}$). After three hours, three photos were taken from each well (3 technical repeats) via an inverted fluorescence microscope (Nikon A1R) and cells were counted using Image-J software.

To assess cell spreading, C2C12 cells were cultured on FCS-treated, and untreated gamma-sterilised hydrogels (n=3). After 24 hours, three photos were taken from each well (3 technical repeats) via inverted fluorescence microscopy and the cell surface area was calculated using Image-J software.

C2C12 cell proliferation was assessed by measuring the cell count in each well on days 3 and 5 via ATPLite assay. These readings were normalised to the reading of cells at day 0 and divided by the total number of days.

*Cell maturation*: 50% confluent C2C12 cells were seeded onto either the tissue culture plastic or gamma-sterilised photo-click hydrogels and continually cultured to 70% confluence (2-3 days). A 4.8-times higher initial cell density (i.e. $4.8\times10^5$ cells·mL$^{-1}$) was applied to the hydrogels with respect to the tissue culture plastic control (i.e. $1\times10^5$ cells·mL$^{-1}$), in line with the decreased cell attachment in the former samples. To induce myotube formation, the media was changed to a myogenic differentiation medium, i.e. DMEM supplemented with 1% PenStrep, 1% FCS and 1% ITS, and replaced every day during cell culture. Three replicates were used for each sample group.

Myoblast maturation assessment was performed by staining C2C12 nuclei (DAPI, Sigma) and F-actin (Oregon Green 488 Phalloidin, Fisher Scientific) at days 3 and 7. The fusion index (*FI*) was calculated from low-magnification fluorescent images by dividing the total number of nuclei in myotubes (≥ 2 nuclei) by the total number of nuclei counted. For samples stained for myosin (MF20, Thermo Fisher Scientific) and cell nucleus, the percentage maturation was calculated by dividing the number of cells stained positive for myosin heavy chain over the total number of cells.

*In vivo study*: All animal studies were conducted under procedures approved by the University of Leeds Ethics Committee and under the UK Home Office project license (PPL: 70/8549). Samples (Ø: 1 cm) of either CollPEG4.5 or Mucograft® (Geistlich) were implanted subcutaneously in six Sprague Dawley (SD) male rats (300-350 g). A pilot study was initially performed to assess hydrogel integrity after 14 days *in vivo*. For the systematic study *in vivo*, the samples were dissected out at days 1, 4 and 7 post-implantation and fixed with 10% neutral buffered formalin (NBF, 24 hours) for histological analysis.

Two rats were used at each time point for each sample. The surgery was carried out under general anaesthesia using trifluorane and 2.5% oxygen on a heated mat. Two full-thickness skin incisions were made on the backs of the rats. Subcutaneous tissues were then blindly dissected using artery forceps to create the sample pockets, and four PBS swollen collPEG4.5 hydrogels were implanted within each pocket (away from the incision). The wounds were closed using 5-0 Ethilion sutures. Four samples of Mucograft® (Geistlich) were used as the commercial control at each time point. The wound was cleaned with sterile water and 75% ethanol, the triflourane was switched off, and the oxygen was left on until the rats were fully recovered. Vertegesic (0.03 mg/mL, 300 μL) injection was given to the rat. At the chosen time point, the rats were sacrificed (schedule 1). The samples and the surrounding tissue were then dissected out and fixed in 10% NBF.

*Histology*: Following 24-hour fixation in 10% NBF, samples were dehydrated, paraffin-embedded, sectioned using a microtome (*h*: 4 μm) and mounted onto glass slides. The samples were deparaffinised with xylene (10 minutes), followed by exposure to a series of gradient ethanol solutions (100%–70%) (5 minutes in each). Samples were subsequently brought to water and incubated in hematoxylin (3 minutes) followed by rinsing in a water bath. Incubation in Scott's tap water (30 seconds) followed prior to staining with eosin (3 minutes). Samples were washed thoroughly in a water bath before being dehydrated to xylene and mounted with a Permount mounting medium.

*Statistical analysis*: Levene's test and t-test were used to test for data variance and for comparison of two different groups, respectively. One-way ANOVA followed by Tukey post-hoc test was performed on data when Levene's test showed $p \geq 0.05$ and equal variances could be assumed. A Welch's ANOVA followed by a Games-Howell post-hoc test was carried out on data when Levene's test showed $p \leq 0.05$, and equal variances could not be assumed. Statistical significance was determined by $p \leq 0.05$. MiniTab software was used for all tests.

**Acknowledgements**

The authors gratefully acknowledge funding from the Engineering and Physical Sciences Research Council (EPSRC) Centre for Doctoral Training in Tissue Engineering and Regenerative Medicine, grant number EP/L014823/1. The authors wish to thank Jackie Hudson and Dr Sarah Myers for technical assistance with imaging and cell culture, respectively. Dr R. Phil. W. Davies and Dr Scott Finlay are also acknowledged for technical assistance with the rheology and compression testing, respectively.